# Coherent control of a V-type three-level system in a single quantum dot

Qu-Quan Wang,[1, 2 *] Andreas Muller,[2] Mu-Tian Cheng,[1] Hui-Jun Zhou,[1]
Pablo Bianucci,[2] and Chih-Kang Shih,[2]

[1] *Department of Physics, Wuhan University, Wuhan 430072, People's Republic of China*
[2] *Department of Physics, University of Texas at Austin, Austin, Texas 78712*
(dated: September 14[th], 2005)

In a semiconductor quantum dot, the $\Pi_x$ and $\Pi_y$ transitions to the polarization eigenstates, $|x\rangle$ and $|y\rangle$, naturally form a three-level V-type system. Using low-temperature polarized photoluminescence spectroscopy, we have investigated the exciton dynamics arising under strong laser excitation. We also explicitly solved the density matrix equations for comparison with the experimental data. The polarization of the exciting field controls the coupling between the otherwise orthogonal states. In particular, when the system is initialized into $|y\rangle$, a polarization-tailored pulse can swap the population into $|x\rangle$, and vice-versa, effectively operating on the exciton spin.

Coherent optical control over individual quantum systems in semiconductors has been the subject of active research over the past decade. It also plays a central role in the current topic of quantum information processing, where quantum bits (qubits) need to be addressed coherently. While excitons confined to semiconductor quantum dots (SQDs) are attractive qubits [1-4], they also provide a fundamental testing ground for coherent light-matter interactions in the solid-state. In the linear excitation regime, the confined exciton's wavefunction can, for instance, be manipulated by tailored pulse-pairs via quantum interference [5]. Under strong field excitation, on the other hand, the upper state of a two-level exciton becomes significantly populated and, with pulse-pair excitation, its dynamics involve the subtle interplay between Rabi oscillations and quantum interference [6,7]. Recently, such capabilities have led to the SQD implementation of the one-qubit Deutsch-Jozsa algorithm [8] as well as the operation of a full two-qubit C-ROT gate [9].

Here we are concerned with non-linear coherent optical control of the fine-structure-split states, $|x\rangle$ and $|y\rangle$, of an excited exciton in a single SQD. These states originate in SQD shape anisotropy [10,11] and play an important role in spin relaxation [12], biexciton creation [9,13], quantum beats and Raman beats [14]. Together with the vacuum state, $|v\rangle$ (no exciton), they naturally define a V-type three-level system, composed of two orthogonal transition dipole moments. In atomic V-type systems, important quantum effects have been pointed out [15,16]. Yet there appear to be few investigations in the solid-state counterpart.

In this letter, we report on photoluminescence (PL) studies of a single self-assembled InGaAs/GaAs SQDs with V-type exciton energy structure. The $\Pi_x$ and $\Pi_y$ transitions to $|x\rangle$ and $|y\rangle$ are excited simultaneously by strong polarization-tailored pulses, resulting in unique dynamics involving the coupled transitions, Rabi oscillations, and quantum interference. In particular, we show that population oscillations between the two orthogonal states are realized, *although a direct transition among them is forbidden*. These interpretations are confirmed by density matrix calculations of the time-evolution of the system.

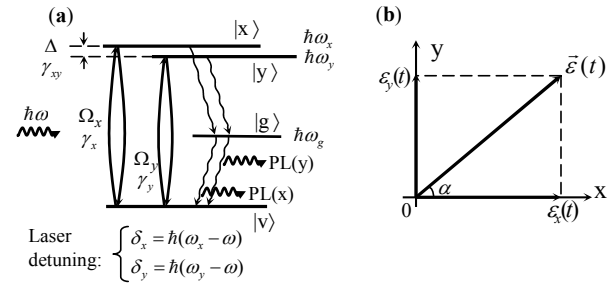

FIG. 1. (a) Schematic SQD energy diagram. The exciton is excited to the first excited state, $|x\rangle$ ($|y\rangle$), then relaxes non-radiatively to the excitonic ground state, $|g\rangle$, and finally radiatively decays back to the vacuum state, $|v\rangle$ (no exciton), emitting x (y)-polarized PL. (b) Laser polarization.

The sample investigated contains $In_{0.5}Ga_{0.5}As$ SQDs grown by molecular beam epitaxy [17]. A mode-locked Ti:Sapphire laser delivering 6 ps long pulses at a repetition rate of 80 MHz is used to excite the sample (maintained at 5 K). The laser is resonant with the transitions from the vacuum state $|v\rangle$ to the first excited states $|x\rangle$ and $|y\rangle$ [Fig. 1(a)]. Measurement of the polarized PL intensity from recombination of ground-state excitons monitors the populations, $\rho_{xx}$ and $\rho_{yy}$, of the excited states |x⟩ and |y⟩ in individual SQDs, i.e. the x-polarized (y-polarized) PL, denoted by PL(x) [PL(y)] is proportional to $\int_0^\infty \rho_{xx} dt$ ($\int_0^\infty \rho_{yy} dt$), when spin-relaxation is negligible. The PL signals were recorded using a spectrometer combined with a 2D liquid nitrogen cooled CCD array detector. The energy splitting, due to the anisotropic electron-hole exchange interaction, was measured from the polarized photoluminescence excitation (PLE) spectrum to be about $\Delta$ =85 μeV for the particular SQD studied, but may vary from dot to dot between ~30 and ~100 μeV. The laser bandwidth, although larger than the |x⟩-|y⟩ energy splitting, is too small to excite a bi-excitonic state. The laser (energy $\hbar\omega$) is also far from resonance with the exciton ground state, |g⟩, whose only role is to monitor the population of the excited states.

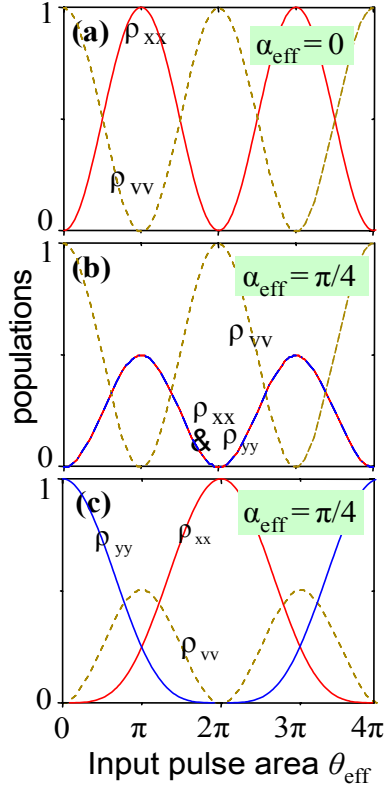

FIG. 2. Theoretical evolution of the three-level V-system as a function of the effective pulse area, for various initial conditions and effective polarization angles.

The matrix elements (dipole approximation) of the interaction Hamiltonian are thus $V_{xg} = \frac{1}{2}\mu_x \varepsilon_x(t) e^{-i\nu t} + c.c.$, $V_{yg} = \frac{1}{2}\mu_y \varepsilon_y(t) e^{-i\nu t} + c.c.$, $V_{xx} = V_{yy} = V_{gg} = 0$, $V_{xy} = V_{yx} = 0$ where $\mu_x$ and $\mu_y$ are dipole moments of |v⟩→|x⟩ and |v⟩→|y⟩ transitions respectively; $\varepsilon_x(t) = \varepsilon_0(t)\cos\alpha$ and $\varepsilon_y(t) = \varepsilon_0(t)\sin\alpha$ are the electric field envelopes along the x and y directions, respectively; and $\alpha$ is the polarization angle [Fig.1(b)].

The dynamics of the system are described using the density matrix formalism. For convenience we define the vector $\vec{S} = (U_1, U_2, U_{xy}, V_1, V_2, V_{xy}, W_1, W_2)$ which contains the "Bloch" vectors $(U_1 = \rho_{xv} e^{i\nu t} + c.c., V_1 = i\rho_{xv} e^{i\nu t} + c.c., W_1 = \rho_{xx} - \rho_{vv})$ and $(U_2 = \rho_{yv} e^{i\nu t} + c.c., V_2 = i\rho_{yv} e^{i\nu t} + c.c., W_2 = \rho_{yy} - \rho_{vv})$ of the |v⟩→|x⟩ and |v⟩→|y⟩ transitions [18], respectively, and $U_{xy} = \rho_{xy} + c.c.$ and $V_{xy} = -i\rho_{xy} + c.c.$. Within the rotating wave approximation, $\vec{S}$ obeys the equation of motion [15,19]

$$\dot{\vec{S}}(t) = M(t)\vec{S}(t) - \Gamma\vec{S}(t) - \vec{\Lambda} \qquad (1)$$

in which

$$\Gamma = \begin{pmatrix} \frac{1}{2}\gamma_x & \frac{1}{2}\gamma_{xy} & 0 & \delta_x & 0 & 0 & 0 & 0 \\ \frac{1}{2}\gamma_{xy} & \frac{1}{2}\gamma_y & 0 & 0 & \delta_y & 0 & 0 & 0 \\ 0 & 0 & \frac{1}{2}(\gamma_x+\gamma_y) & 0 & 0 & -\Delta & \frac{1}{3}\gamma_{xy} & \frac{1}{3}\gamma_{xy} \\ -\delta_x & 0 & 0 & \frac{1}{2}\gamma_x & \frac{1}{2}\gamma_{xy} & 0 & 0 & 0 \\ 0 & -\delta_y & 0 & \frac{1}{2}\gamma_{xy} & \frac{1}{2}\gamma_y & 0 & 0 & 0 \\ 0 & 0 & \Delta & 0 & 0 & \frac{1}{2}(\gamma_x+\gamma_y) & 0 & 0 \\ 0 & 0 & \frac{1}{2}\gamma_{xy} & 0 & 0 & 0 & 0 & 0 \\ 0 & 0 & \frac{1}{2}\gamma_{xy} & 0 & 0 & 0 & 0 & 0 \end{pmatrix} \qquad (2)$$

and $\vec{\Lambda} = (0,0,\frac{2}{3}\gamma_{xy},0,0,0,\frac{1}{3}\gamma_x,\frac{1}{3}\gamma_y)^T$ account for the various decay rates and the detuning [Fig. 1(a)]. Since we are concerned with the time-evolution of $\vec{S}$ caused by tailored laser pulses, we consider the general situation when the SQD is excited by a pair of pulses with mutual phase delay $\phi=2\pi\nu t_d$. $M(t)$ then reads:

$$M = \frac{1}{2}\begin{pmatrix}
0 & 0 & -\Omega_{y2}\sin\phi & 0 & 0 & -(\Omega_{y1}+\Omega_{y2}\cos\phi) & -2\Omega_{y2}\sin\phi & 0 \\
0 & 0 & -\Omega_{x2}\sin\phi & 0 & 0 & \Omega_{x1}+\Omega_{x2}\cos\phi & 0 & -2\Omega_{y2}\sin\phi \\
\Omega_{y2}\sin\phi & \Omega_{x2}\sin\phi & 0 & \Omega_{y1}+\Omega_{y2}\cos\phi & \Omega_{x1}+\Omega_{x2}\cos\phi & 0 & 0 & 0 \\
0 & 0 & -(\Omega_{y1}+\Omega_{y2}\cos\phi) & 0 & 0 & \Omega_{y2}\sin\phi & -2(\Omega_{x1}+\Omega_{x2}\cos\phi) & 0 \\
0 & 0 & -(\Omega_{x1}+\Omega_{x2}\cos\phi) & 0 & 0 & -\Omega_{x2}\sin\phi & 0 & -2(\Omega_{y1}+\Omega_{y2}\cos\phi) \\
\Omega_{y1}+\Omega_{y2}\cos\phi & -(\Omega_{x1}+\Omega_{x2}\cos\phi) & 0 & -\Omega_{y2}\sin\phi & \Omega_{x2}\sin\phi & 0 & 0 & 0 \\
2\Omega_{x2}\sin\phi & \Omega_{y2}\sin\phi & 0 & 2(\Omega_{x1}+\Omega_{x2}\cos\phi) & \Omega_{y1}+\Omega_{y2}\cos\phi & 0 & 0 & 0 \\
\Omega_{x2}\sin\phi & 2\Omega_{y2}\sin\phi & 0 & \Omega_{x1}+\Omega_{x2}\cos\phi & 2(\Omega_{y1}+\Omega_{y2}\cos\phi) & 0 & 0 & 0
\end{pmatrix} \quad (3)$$

For laser pulses with hyperbolic secant time profile, the instantaneous Rabi frequencies $\Omega_{x1} = (\mu_x/\hbar)\cos\alpha_1 \varepsilon_{01} Sech((t-t_0)/\tau_p)$ and $\Omega_{y1} = (\mu_y/\hbar)\sin\alpha_1 \varepsilon_{01} Sech((t-t_0)/\tau_p)$ describe the interaction of the first pulse with the $|v\rangle \rightarrow |x\rangle$ and $|v\rangle \rightarrow |y\rangle$ transitions, respectively, while $\Omega_{x2} = (\mu_x/\hbar)\cos\alpha_2 \varepsilon_{02} Sech((t-t_0-t_d)/\tau_p)$ and $\Omega_{y2} = (\mu_y/\hbar)\sin\alpha_2 \varepsilon_{02} Sech((t-t_0-t_d)/\tau_p)$ account for the interaction of the second pulse with the same transitions. The temporal width $\tau_p$ of the pulses is maintained constant but the polarization angles $\alpha_1$ and $\alpha_2$ of the two pulses are variable.

Under single-pulse excitation, i.e. when $\Omega_{x2} = \Omega_{y2} = 0$, a simple analytical solution exists, assuming no decoherence and no detuning ($\Gamma=0$, $\bar{\Lambda}=0$). It is convenient to define the effective polarization angle $\alpha_{eff} = \arctan(\mu_y \sin\alpha / \mu_x \cos\alpha)$, effective input pulse area $\theta_{eff}(t)$ and effective transition dipole moment $\mu_{eff}$,

$$\theta_{eff}(t) = (\mu_{eff}/\hbar)\int_{-\infty}^{t}\varepsilon(t')dt' \quad (4)$$

$$\mu_{eff} = \sqrt{\mu_x^2 \cos^2\alpha + \mu_y^2 \sin^2\alpha} \quad (5)$$

The solutions of Eq.(1), in terms of the populations of $|x\rangle$ and $|y\rangle$ then read:

$$\begin{cases} \rho_{yy} = \sin^2\alpha_{eff}\sin^2(\tfrac{1}{2}\theta_{eff}), & \rho_{xx}(0)=\rho_{yy}(0)=0; \\ \rho_{xx} = \cos^2\alpha_{eff}\sin^2(\tfrac{1}{2}\theta_{eff}), & \rho_{xx}(0)=\rho_{yy}(0)=0; \end{cases} \quad (6)$$

$$\begin{cases} \rho_{yy} = \sin^2(2\alpha_{eff})\sin^4(\tfrac{1}{4}\theta_{eff}), & \rho_{xx}(0)=1; \\ \rho_{xx} = [1 - 2\cos^2\alpha_{eff}\sin^2(\tfrac{1}{4}\theta_{eff})]^2, & \rho_{xx}(0)=1; \end{cases} \quad (7)$$

Furthermore, if $\alpha_{eff} = \pi/4$, the population difference between the two sub-states has the general form,

$$\rho_{yy} - \rho_{xx} = (\rho_{yy}(t_0) - \rho_{xx}(t_0))\cos(\tfrac{1}{2}\theta_{eff}) \quad (8)$$

Equations (6), (7), and (8) reveal three interesting characteristics of the population oscillation as a function of $\theta_{eff}$, in this system: (i) the populations of $|x\rangle$ and $|y\rangle$ oscillate with the *same* period (i.e. Rabi frequency) even though $\mu_x \neq \mu_y$. (ii) The effective transition dipole moment $\mu_{eff}$ of $|v\rangle \rightarrow |xy\rangle$ is *tunable* in the range $[\mu_x, \mu_y]$, with $|xy\rangle = a|x\rangle + b|y\rangle$. (iii) The period of the population oscillations is $2\pi$ when $\rho_{xx}(0) = 0$ and is $4\pi$ when $\rho_{xx}(0) = 1$. The population oscillations of $|x\rangle$, $|y\rangle$, and $|v\rangle$ are depicted in Fig. 2 as a function of $\theta_{eff}$ for various initial conditions and polarization angles. When $(\rho_{yy}(0) = \rho_{xx}(0) = 0, \alpha_{eff} = 0)$ [Fig. 2(a)], the excitation field only couples to the $|v\rangle \rightarrow |x\rangle$ transition and the system undergoes the familiar two-level Rabi oscillations. On the other hand, when $(\rho_{yy}(0) = \rho_{xx}(0) = 0, \alpha_{eff} = \pi/4)$ [Fig. 2(b)], the two transitions are coupled, yet the populations of $|x\rangle$ and $|y\rangle$ simultaneously undergo oscillations. Finally, under asymmetrical initial conditions $(\rho_{yy}(0) = 1, \alpha_{eff} = \pi/4)$ [Fig. 2(c)], coherent population flopping between $|x\rangle$ and $|y\rangle$ occurs, even though $\langle x|y\rangle = 0$.

In practice, these results are quantitatively affected by decoherence, inevitably present in the system. For experimental comparison we thus have to resort to numerical integration of Eq. (1) using the two-pulse matrix $M(t)$ in which the first pulse acts as an initialization pulse for the case $\rho_{xx}(0)$, $\rho_{yy}(0) \neq 0$. Dephasing prevents initialization into a stationary state, and instead brings the system into a superposition state which is allowed to freely evolve in the non-rotating frame. Therefore, the relative phase between the two pulses plays an important role and is reflected in the data by fine-time PL oscillations.

The polarized PL signals are shown in Fig. 3 as a function of input pulse area without any pre-pulse [Fig. 3(a)(b)(c)], and with a y-polarized $\pi$-*pre-pulse* [Fig. 3(d)(e)(f)]. In both cases, the polarization of the manipulation pulse was fixed at $\alpha = \pi/4$ while its pulse area was varied by changing the laser intensity.

In the absence of a pre-pulse, both PL(x) and PL(y) oscillate with the same period as expected for $\rho_{xx}$ and $\rho_{yy}$ [Eq. (6)], and their difference, PL(x)-PL(y) almost vanishes [Fig. 3(c)].

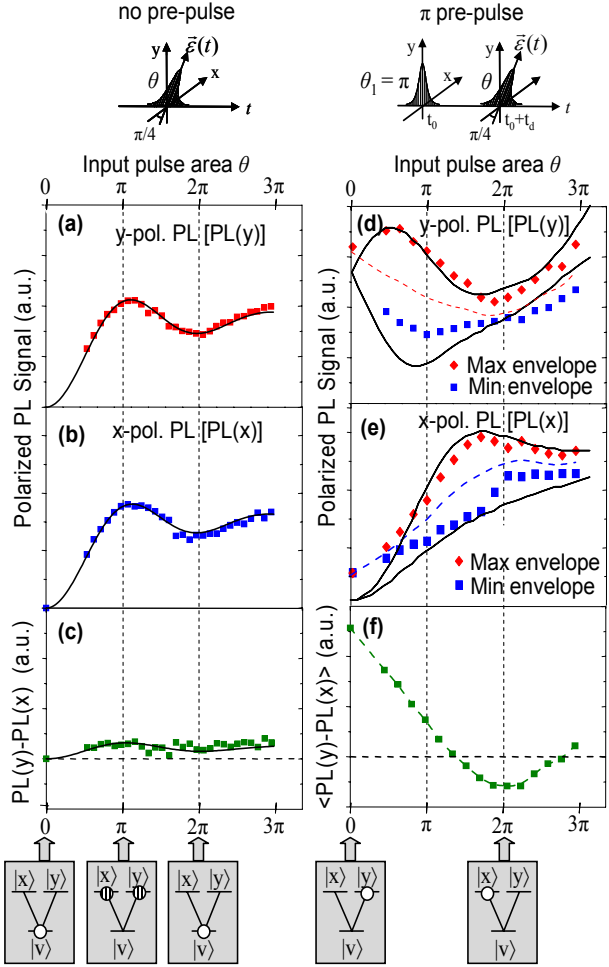

FIG. 3. Experimental evolution of the three-level V-system, using single pulse (left) and two-pulse excitation (right). y-polarized (a) and x-polarized (b) PL signal as a function of pulse area, for an excitation polarization angle $\alpha = \pi/4$. The difference PL(y)-PL(x) is plotted in (c). y-polarized (d) and x-polarized (e) PL signal as a function of pulse area, using a $\pi$-pre-pulse with $\alpha = \pi/2$. The temporal delay between the two pulses was fixed while the pulse area of the second pulse with $\alpha = \pi/4$ was varied. For each value of the pulse area, the relative phase between the two pulses was scanned over one period and the PL maxima and minima were recorded. The solid lines represent the simulated envelopes. (f) Difference between the averages of PL(x) and PL(y). The schematics at the bottom are meant to illustrate the population floppings pictorially.

The latter behavior validates the assumption that spin-relaxation is negligible during the carrier relaxation from the exciton excited state to the exciton ground state. The population oscillations are also strongly damped due to dephasing processes which have been included in the simulated oscillations [solid curves in Fig. 3(a)(b)], obtained from numerical integration of Eq.(1). The source of the oscillation damping most likely originates in transitions to and from the wetting layer, whose formation can potentially be suppressed [20].

Using a y-polarized $\pi$-pre-pulse, the initial condition $\rho_{yy}(0) = 1$ can be simulated. The second pulse ($\alpha = \pi/4$) then induces population dynamics in the system governed by $M(t)$ [Eq.(3)], leading to the phase-sensitive evolution of the polarized PL [Fig. 2(d)(e)]. The coarse delay between the pulses was fixed at 12 ps to prevent mutual temporal overlap while their phase delay was varied with a piezo-controlled fine-time delay and the maxima (diamonds) and minima (squares) of PL(y) and PL(x) recorded [Fig. 3(d) and Fig. 3(e), respectively]. The *phase-averaged* values <PL(x)> and <PL(y)> are also plotted [dashed curves in Fig. 3(d)(e)]. These represent the general trend for the population transfer, namely the population of state |y⟩ decreases at the expense of |x⟩, corresponding to the population swapping without direct transition described by Eq. (8) and Fig. 2(c). Numerical integration of Eq. (1), including dephasing, reproduces the envelopes of the oscillations reasonably well [solid curves in Fig. 3(e)(f)].

The complex dynamics of this population transfer has its roots in the coupling between the $\Pi_x$ and $\Pi_y$ transitions via the common ground state |v⟩. The strength of this coupling can be represented by the normalized quantity $f_c = 1-(W_{1,\,max}-W_{2,\,max})/(W_{1,\,max}+W_{2,\,max})$. There is no coupling ($f_c=0$) when $\alpha_{eff}= 0$ or $\pi/2$, in which case the V-type three-level system reduces to a two-level system. The coupling reaches a maximum ($f_c=1$) when $\alpha_{eff} = \pi/4$. The difference of the third component of the two coupled optical Bloch vectors will then oscillate in the form $W_1(t)-W_2(t)=(W_1(t_0)-W_2(t_0))\cos(\theta_{eff}/2)$ [Eq. (8)], with the initial value set by the polarized pre-pulse. Indeed, this is what is observed experimentally and reflected in the difference <PL(y)-PL(x)> between the polarized PL signals [Fig. 2(f)].

As is well-known in SQDs, the anisotropic exchange interaction is responsible for the splitting of the originally degenerate heavy-hole exciton spin states $|m_h = 3/2\rangle$ and $|m_h = -3/2\rangle$, into the states $|x\rangle = (|m_h = 3/2\rangle + |m_h = -3/2\rangle)/\sqrt{2}$, and $|y\rangle = (|m_h = 3/2\rangle - |m_h = -3/2\rangle)/\sqrt{2}$ [10]. The coherent manipulation of the populations of $|x\rangle$ and $|y\rangle$ with tailored pulses is thus equivalent to a manipulation of the exciton's spin state, with control over both phase and amplitude of the quantum states.

In conclusion, we have examined both theoretically and experimentally the exciton dynamics arising under polarization-tailored two-pulse excitation of the V-system defined by the polarization eigenstates $|x\rangle$, $|y\rangle$ and the crystal ground state $|v\rangle$. This system is special in that the coupling between the otherwise orthogonal states $|x\rangle$ and $|y\rangle$ can be polarization-tuned. Although dephasing affects the detailed time-evolution, the essential characteristics are captured by the analytic solution to the density matrix equations with $\Gamma, \bar{\Lambda} = 0$. The capabilities demonstrated here present an additional step towards all-optical non-linear coherent control of a multilevel excitonic system.

We gratefully thank Prof. L. J. Sham, Prof. R. B. Liu, Prof. C. Piermarocchi and Prof. Q. K. Xue for their comments and suggestions. This work was supported by NSFC (10344002 and 10474075), NSF (DMR-0210383 and DMR-0306239), the Office of Naval Research (N0014-04-1-0336), the Texas Advanced Technology program and the W. M. Keck Foundation.

* Corresponding author electronic address:
   qqwang@physics.utexas.edu